# Dispersion Studies on Multimode Polymer Spiral Waveguides for Board-Level Optical Interconnects

**Jian Chen[1], Nikos Bamiedakis[1], Tom J. Edwards[2], Christian T.A. Brown[2], Richard V. Penty[1], and Ian H. White[1]**

[1]Electrical Engineering Division, Department of Engineering, University of Cambridge, 9 JJ Thomson Avenue, Cambridge, CB3 0FA, UK
[2]SUPA, School of Physics & Astronomy, University of St Andrews, North Haugh, St Andrews, Fife, KY16 9SS, UK
Author e-mail address: jc791@cam.ac.uk

**Abstract:** Dispersion studies are conducted on 1m long multimode polymer spiral waveguides with different refractive index profiles. Bandwidth-length products >40GHz×m are obtained from such waveguides under a 50/125 μm MMF, indicating the potential of this technology.

## 1. Introduction

In supercomputer and data centre environments, short-reach optical interconnects have attracted significant interest in recent years owing to their advantages over conventional copper-based interconnects, such as higher bandwidth, immunity to electromagnetic interference and reduced power consumption [1, 2]. Multimode polymer waveguides, in particular, are an attractive technology for use in cost-efficient board-level optical interconnects. They can be directly integrated on printed circuit boards (PCBs) owing to their favourable material properties and enable low-cost system assembly with relaxed alignment tolerances owing to their large core dimensions (typically 30 – 70 μm). Their highly-multimoded nature however, raises important questions about their potential to support very high data rates. This concern regarding the bandwidth performance of such waveguides becomes particular important given the continuous improvement in high-speed performance of vertical-cavity surface-emitting lasers (VCSELs). Directly-modulated 850 nm VCSELs operating at 64 Gb/s have been recently demonstrated [3]. We have recently presented thorough bandwidth studies on a 1 m long polymer multimode spiral waveguide using frequency domain ($S_{21}$) measurements, and we have demonstrated a bandwidth-length product of at least 35 GHz×m [4]. However, we haven't been able to determine the actual bandwidth of these waveguides due to the limitations in the active components and instruments. In this paper, we present studies on the bandwidth performance of 1 m long multimode polymer spiral waveguides using pulse broadening measurements. Two waveguide samples with different refractive index profiles are tested and bandwidth-length products in excess of 40 GHz×m and 70 GHz×m are demonstrated under a 50/125 μm MMF launch. The results highlight the capabilities of this waveguide technology and indicate the potential of achieving data transmission of 100 Gb/s over a single waveguide.

## 2. Waveguide samples and experimental set-up

Short pulse measurements are conducted on 1 m long spiral multimode polymer waveguides. The waveguides are fabricated on 8-inch silicon substrates with conventional photolithography from siloxane polymer materials (core/cladding: Dow Corning® WG-1020/ XX-1023 Cured Optical Elastomer). The refractive index profile of the waveguides can be adjusted by controlling the fabrication parameters [5]. For these experiments, two waveguide samples with a similar spiral pattern and length but with different refractive index profiles are employed. One sample is produced so that it exhibits a step-like index (SI) profile, while the other has a more gradual graded-index (GI) profile. The actual refractive index profiles of the samples are currently being measured and will be presented at the conference. The two samples have similar core sizes ~ 32×35 μm$^2$, while the waveguide length is 105.5 cm. Images of the spiral structure and waveguide output facets are shown in Fig. 1a-c.

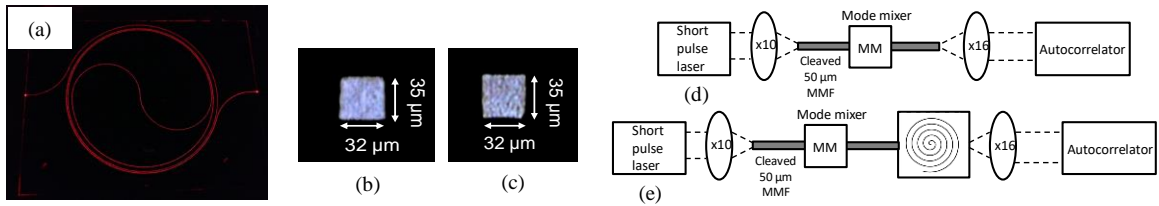

Fig. 1. Images of (a) the 1 m long spiral waveguide illuminated with red light and facet of the (b) SI and (c) GI waveguide illuminated with 850 nm light. (d-e) Experimental setup for the (d) back-to-back and (e) waveguide link using a 50/125 μm MMF launch and the mode mixer.

The experimental setup used for the measurement of the transmitted pulse width with and without (back-to-back) the polymer waveguide is shown in Fig. 1d and 1e respectively. A Ti:Sapphire laser operating at 850 nm is used as the source while a FR103-MN autocorrelator is used at the receiver end. A short cleaved 50/125 μm MMF patchcord is employed to couple the light into the waveguide as such launch is highly likely in real-world systems. The light emitted from the laser is coupled into the 50/125 μm MMF patchcord with a 10× microscope objective (NA=0.25) while a mode mixer (MM, Newport FM-1) is used to generate a more uniform power distribution inside the input fibre. The cleaved end of the input MMF is placed on a precision translational stage and a displacement sensor is used to control the launch position in the horizontal direction. Two different launch conditions are studied, with (MM) and without the MM (no-MM) in order to investigate the waveguide performance, as different mode power distributions at the waveguide input result in different levels of modal dispersion at the waveguide output. The output light from the waveguide is collected with a 16× microscope objective (NA=0.32) and is coupled into the autocorrelator. For each one of the two launches studied, the received optical power and transmitted pulse are recorded for different positions of the input fibre.

## 3. Experimental results and discussion

The obtained autocorrelation traces for the waveguide and back-to-back case are converted to signal pulses using curve fitting and common pulse shapes (e.g. sech$^2$, Gaussian). The frequency response of the waveguide and back-to-back link are calculated by taking the Fourier transform of the signal pulses, and the frequency response of the waveguide is obtained by subtracting the back-to-back link frequency response from the waveguide link response. The 3dB bandwidth for each waveguide sample is found for each input position and for each launch, and is plotted in Fig. 2 as a function of the horizontal offset along with the respective normalised received power.

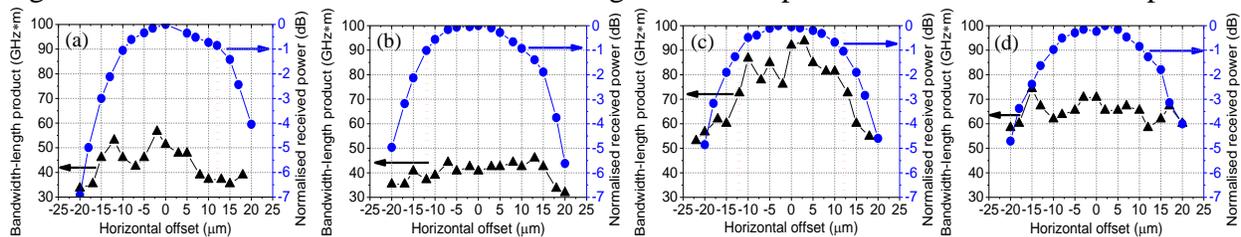

Fig. 2 Calculated 3dB bandwidth (black triangle) and measured normalised received power (blue circle) of the 1 m long spiral polymer samples (a-b) SI waveguide under a (a) no-MM and (b) MM launch; (c-d) GI waveguide under a (c) no-MM and (d) MM launch.

The results demonstrate that, as expected the waveguide with the GI profile exhibits higher bandwidth than the SI waveguide. Moreover, the use of the mode mixer at the waveguide input results in a reduced bandwidth performance. Bandwidth-length products > 40 GHz×m and 70 GHz×m are observed within the 1dB alignment tolerances for the SI and GI waveguides (~ ±10 µm, noted with red lines in plots), respectively under the 50/125 µm MMF launch without the MM. The use of the MM results in lower bandwidth values but in a smaller variation across the input offsets: ~40 GHz×m for the SI sample and ~65 GHz×m for the GI sample. The large bandwidth values obtained from the samples can be attributed to the spiral structure of the waveguide which results in larger attenuation for the higher order modes and therefore, in reduced multimode dispersion in the guides. The large values obtained (in excess of 70 GHz) for the GI sample demonstrate that sufficient bandwidth to support 100 Gb/s data transmission over a single waveguide channel can be achieved through refractive index engineering in such waveguides. A simulation model taking into account the waveguide index profile, the mode loss profile and input launch condition is being developed in order to match the experimental results and detailed bandwidth simulation studies will be presented at the conference.

## 4. Conclusions

Pulse broadening measurements are conducted on 1 m long polymer multimode waveguide samples with different refractive index profiles. Bandwidth-length products > 40 GHz×m and 70 GHz×m are demonstrated for SI and GI waveguide samples respectively under a 50/125 µm MMF launch, indicating the capability of 100 Gb/s data transmission over a single multimode waveguide channel with appropriate engineering of refractive index.

## 5. Acknowledgements

The authors would like to acknowledge Dow Corning for providing the waveguide samples and EPSRC for supporting the work. Data related to this publication is available at the University of Cambridge data repository.